\documentclass[preprint,12pt]{elsarticle}
\usepackage{times,fullpage}
\usepackage{url,color}
\usepackage {cite}
\usepackage{graphicx}
\usepackage[bf]{subfigure}
\begin{document}

\newcommand{\rednote}[1]{\textcolor{red}{ \em #1 }}


\begin{frontmatter}
\title{Monitoring wild animal communities with arrays of motion sensitive camera traps\tnoteref{t1}}

\author[nym,sri]{Roland Kays}
\ead{rkays@mail.nysed.gov}

\author[ucsd]{Sameer Tilak}
\ead{stilak@ucsd.edu}

\author[mpi,ug]{Bart Kranstauber}
\ead{kranstauber@mail.orn.mpg.de}

\author[sri,wu,ug]{Patrick A. Jansen}
\ead{patrick.jansen@wur.nl} 

\author [zsl]{Chris Carbone}
\ead{chris.carbone@ioz.ac.uk} 

\author[zsl]{Marcus J. Rowcliffe}
\ead{Marcus.Rowcliffe@ioz.ac.uk}
 
\author[ucsd]{Tony Fountain}
\ead{tfountain@ucsd.edu} 

\author[mu]{Jay Eggert}
\ead{jayeggert@mail.mizzou.edu} 

\author[mu]{Zhihai He}
\ead{hezhi@missouri.edu} 

\address[nym]{New York State Museum}
\address [sri]{Smithsonian Tropical Research Institute}
\address[mpi]{Max Planck Institute for Ornithology}
\address[zsl]{Institute of Zoology, Zoological Society of London}
\address[ucsd]{University of California at San Diego	}

\address[mu]{University of Missouri}

\address[ug]{University of Groningen}

\address[wu]{Wageningen University}

\tnotetext[t1]{This work is partially supported by a grant from the NSF DBI 0756920 and NSF Bio 0717071. This paper is significantly extends the research published in the following workshop paper: Roland Kays, Bart Kranstauber, Patrick A. Jansen, Chris Carbone, Marcus Rowcliffe, Tony Fountain, and Sameer Tilak, "Camera traps as sensor networks for monitoring animal communities," IEEE International Workshop on Practical Issues in Building Sensor Network Applications held in conjunction with the 34th annual IEEE Conference on Local Computer Networks (LCN), 2009.}


\begin{abstract}

Studying animal movement and distribution is of critical importance to addressing environmental challenges including invasive species, infectious diseases, climate and land-use change. Motion sensitive camera traps offer a visual sensor to record the presence of a broad range of species providing location Ðspecific information on movement and behavior. Modern digital camera traps that record video present new analytical opportunities, but also new data management challenges. This paper describes our experience with a terrestrial animal monitoring system at Barro Colorado Island, Panama. Our camera network captured the spatio-temporal dynamics of terrestrial bird and mammal activity at the site - data relevant to immediate science questions, and long-term conservation issues. We believe that the experience gained and lessons learned during our year long deployment and testing of the camera traps as well as the developed solutions are applicable to broader sensor network applications and are valuable for the advancement of the sensor network research. We suggest that the continued development of these hardware, software, and analytical tools, in concert, offer an exciting sensor-network solution to monitoring of animal populations which could realistically scale over larger areas and time spans.

\end{abstract}


\end{frontmatter}

\section{Introduction}\label{sec:introduction}

Sensor networks represent a new paradigm for reliable environment monitoring and information collection~\citep{Cerpa01, Dinh07, Gilman05, Guillermo08, Lynette06, Mainwaring02, Selavo07, Szewczyk04, Szlaveczcorr07}. They hold the promise of revolutionizing sensing in a wide range of application domains because of their reliability, accuracy, flexibility, cost-effectiveness, and ease of deployment. A large number of numbers of sensors are being embedded in the natural environments (lakes, rivers, oceans, forest canopy etc.) and civil infrastructure (buildings and bridges, data centers, etc.)~\citep{Holger-07, Kahn-99, Krishnamachari-05}. These sensors produce huge volumes of data that must be acquired, transported, stored, analyzed, and visualized to gain unprecedented scientific and engineering insight. Sensor networks will be at the heart of the next century discovery and are therefore linchpin of e-Science. In this paper, we describe our experiences deploying and testing a real-world sensor network for monitoring animal communities.
 
The movement of organisms through their environment lies at the heart of ecological field research 
and is of critical importance to addressing environmental challenges including invasive species, infectious diseases, climate and land-use change~\citep{Nathan-09}. Movement is the key defining character of most animals, and there are two basic ways to record animal motion~\citep{Turchin98}. The Lagrangian approach monitors a specific individual, for example with a GPS-tag, and records a series of locations it passes through. The Eulerian approach, on the other hand, monitors a specific location and records the movement of all organisms across it. Animal trackers  following the Lagrangian approach have been tracking animal movement since the advent of radio-telemetry~\citep{Lord62}. While useful for many purposes, these individual tracking studies are limited  by the difficulty and bias associated with capturing animals, as well as the logistical complications of tracking over long periods or large areas. Camera traps offer a Eulerian solution to monitoring animals that avoid these biases by simply recording a photograph of the animals that move in front of them. 

Distributed, motion-sensitive cameras (aka camera traps) are examples of sensor networks that can collect data on animal populations. This paper develops the concept of camera traps as a network of distributed sensors to monitor animal communities, using a year-long case study from Barro Colorado Island (BCI), Panama. The developed system uses existing camera technology to capture a unique and unbiased picture of environmental dynamics for medium and large sized terrestrial animals. 
In the remainder of this section we present a background on the use of camera traps and describe the specific study objectives. In Section~\ref{sec:infrastructure} we describe the overall hardware and software infrastructure followed by study design and methodology in Section~\ref{sec:procedure}. In Section~\ref{sec:stat} we briefly describe how camera trap data can be analyzed to answer important biological questions such as diversity and abundance of species. In Sections~\ref{sec:results} we describe experiments results obtained from our year long deployment at BCI. In Section~\ref{sec:pract} we describe several practical aspects of deploying and testing a real-world camera networks. In Section~\ref{sec:futurework} we describe future work and in Section~\ref{sec:conclusion} we present concluding remarks. 



\begin{figure*}
\begin{center}
\mbox{
\subfigure[An map of the camera trap BCI deployment.\label{fig:bci-biased-ct}]{\includegraphics[scale=0.25]{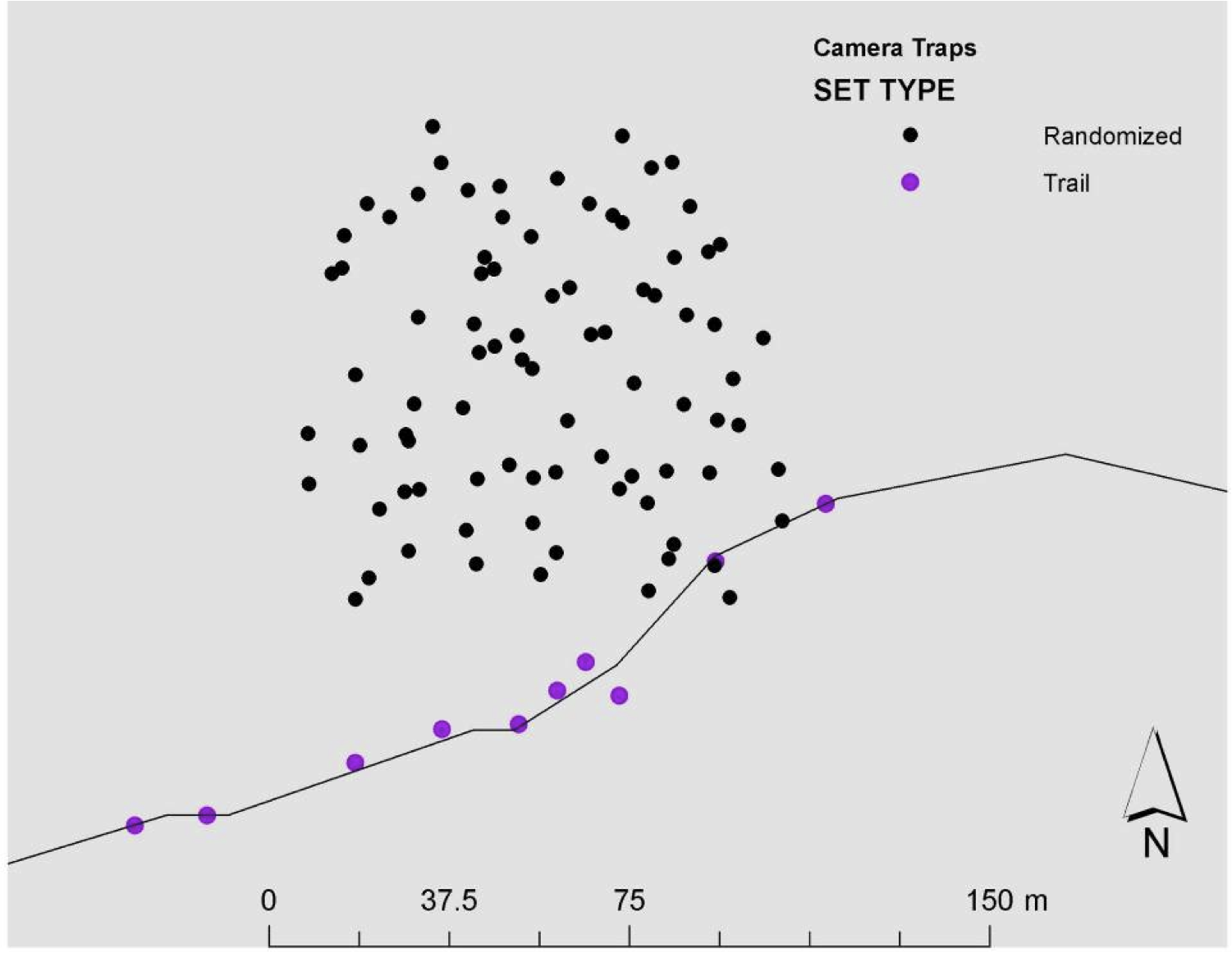}} \quad
\subfigure[A sample camera trap deployment at BCI.\label{fig:bci-camera}]{\includegraphics[scale=0.45]{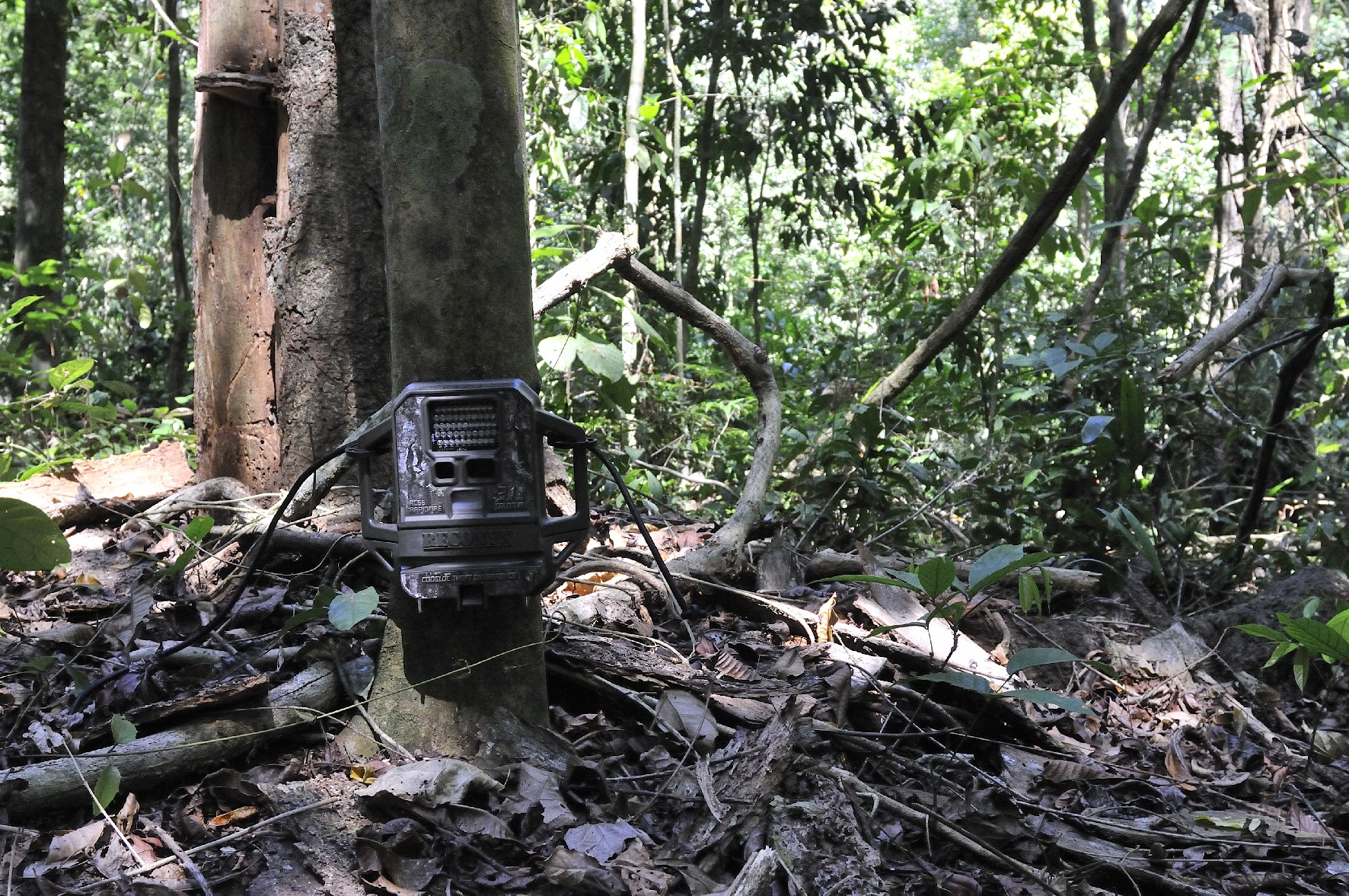}}
}
\caption{Camera Trap Deployment at BCI}
\label{fig:power-study}
\end{center}
\end{figure*}


\subsection{Basic Advantage of Cameratraps}

All animals move, but most are shy and quiet. Camera traps are an
appropriate technique for animal monitoring~\citep{Kays08} for the following
reasons: (1) They are non-invasive: when photographs are captured using invisible IR flashes, camera traps have no effect on most animal behavior. (2) They require low labor: camera traps are easy to deploy and can function for weeks with no attention  (3) They yield robust data: photographs are analogous to museum specimens in
being a permanent record of date, location, and species. (4) They produce bonus material: in addition to recording the presence of a species camera traps can record animal behavior which can be important for scientific questions, but also offers engaging images useful for education and promotion.

\subsection{General Scientific Uses for Camera Trap Data}\label{subsec:ct-generic}

At the most basic level, camera trap data can be used to prove the existence of a species at a site; with sufficient effort, it can also suggest the absence of a species~\citep{MacKenzie-02}.  This can be important to show the arrival of an invasive species, or document the conservation status of rare species~\citep{Pitra06, Cardoza02}.  Multiple georeferenced locations for a species can further be used to document their distribution in an area, and compare with environmental features to create models of distribution or resource selection~\citep{Zielinski05}.  Local animal density, the gold standard for animal monitoring, can also be estimated from camera trap data, given proper study design~\citep{Rowcliffe-08,Karanth98}.  These data become more valuable as they accumulate across sites or over years, for example showing predator-prey relationships of tigers across India~\citep{Karanth04} and documenting their population demography at one site for 9 years~\citep{Karanth06}. 

\begin{figure*}
\begin{center}
\mbox{
\subfigure[Database schema for the camera trap study\label{fig:bci-dbschema}]{\includegraphics[scale=0.35]{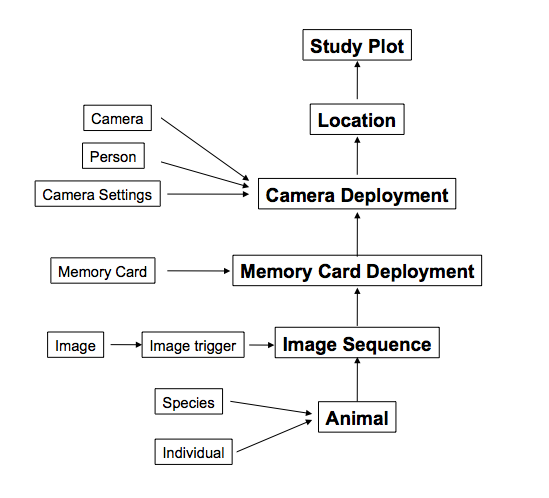}} \quad
\subfigure[Sample clip for image analysis.\label{fig:bci-clip}]{\includegraphics[scale=0.25]{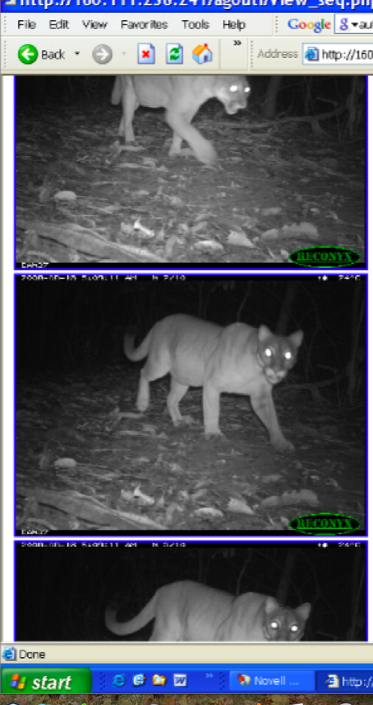}}
}
\caption{Current Data Management Infrastructure and Image Analysis Techniques at BCI.}
\label{fig:power-study}
\end{center}
\end{figure*}    

\subsection{Specific Objectives of our Camera Trap Study}

We used camera traps to survey the diversity and abundance of the terrestrial mammal and bird  communities on BCI. In addition to the general objectives (mentioned in Section~\ref{subsec:ct-generic}), we are also interested to determine how these varied in space with the abundance of keystone fruit resources.
Using aerial photographs and on-the-ground mapping of palm trees~\citep{Jansen08} we have identified 5 low-fruit and 5 high-fruit 1-ha plots. We are comparing levels of animal diversity, activity and abundance in these plots using cameras deployed in random locations within the plots.  We use 2 cameras per plot, moving them to new locations
each 8 days. We followed this protocol for 1 full year from 22 January 2008 to 21 January 2009. We are also studying other things about these plots, chief among them, radio-tracking agoutis and the seeds they disperse. In addition, this experimental setup allows us to look at the effect of food abundance on animal communities.
  
\subsection{Novel Aspects of Our Study}

 Over last few years, wireless sensor networks have been used extensively for ecological monitoring applications~\citep{Dinh07, Gilman05, Guillermo08, Hartung06, Mainwaring02, Selavo07, Szewczyk04}. However, to the best of our knowledge this is the first year-long camera trap deployment in a real-world setting (tropical rain forest on BCI) that uses novel camera deployment strategies and systematically reports back practical and theoretical lessons learned, both from science and sensor network research viewpoint. Following are the key differences between our work and the existing research.

Traditional camera trap studies used film cameras to study one particular target species.  This led to the development of techniques that maximize their efficiency of photographing that species, but may decrease the detection of others (e.g. using baits or targeting animal trails).    Our study aimed to document the entire terrestrial mammal community, and therefore modified protocols to minimize bias and detect any and all animals passing in front of a camera's sensor.
  Four aspects of our protocols are therefore different than most other camera studies: randomizing camera deployment locations, using no bait, monitoring year-round, and recording video sequences for each trigger.  

Additionally, our study was designed to focus camera deployments within 10 study plots to compare animal communities between sites with different amounts of fruit.  More general monitoring protocols would probably alter this slightly to spread the cameras
out more. The year-round monitoring may be overkill for some research objectives.  However, we advocate that randomizing camera locations and setting without bait are important protocols that should be employed by any study trying to document entire animal communities.

\section{Infrastructure}\label{sec:infrastructure}

The field work was conducted at the Barro Colorado Island (BCI) ($9^{o}10^{Õ'} N, 79^{o}51^{Õ'} W$) research station, that is managed by the Smithsonian Tropical Research Institute. BCI is a completely forested, 1567-ha island that was formed when Lake Gatun was created as part of the Panama Canal. Animals continue to move between the island and the surrounding National Park land, which are separated by a few 100m at various places. The island receives an average of 2632 mm of rain per year. The meteorological year is divided into two parts: a pronounced dry season (approximately from mid-December to the end of April), and a wet season (May to mid- December). 
Relative humidity, soil moisture, air pressure, solar radiation, evapotranspiration, wind speed and direction all show marked seasonal variation (wet/dry season differences). On the other hand, temperature varies relatively little throughout the year~\citep{Leigh99}.

\subsection{Camera Hardware Requirements:}

The components of a camera trap sensor network are simple in being; a collection of camera traps which are deployed in the field, a series of memory cards used to record images and transfer them back to the lab, and a database to save and organize images and metadata.  
Live transmitting of data is limited by the battery power needed to send thousands of images from a remote camera, not to mention limited communication networks in many wild areas.  

Camera trap studies do not typically require high-resolution images, but do have a number of minimum requirements (Table~\ref{tab:cam-req}) needed to collect robust and unbiased data .  Because they are typically deployed for long periods of time in harsh conditions, they must be incredibly well protected from rain and humidity (e.g., BCI is a rainforest with a prominent wet season). Most modern digital cameras can capture night-time images using IR flashes, which can not be seen by animals.  This is an important feature because a visible flash is a source of potential bias for a camera trap study if animals are disturbed by the flash and avoid the camera thereafter~\citep{Wegge04}. Digital cameras with infrared flashes should result in neither aversion nor curiosity, although their flashes may still be visible by people if viewed directly.

\begin{table*}
\caption{Hardware requirements for our application of remote cameras as sensor network to monitor animal populations.}
\begin{tabular}{|c|c|}\hline
\textbf{Specification} & \textbf{Requirements}  \\ \hline
Motion Sensor & 5-10m range \\ \hline
Flash & Infrared  \\ \hline
Camera & IR sensitive camera for night pictures, color for day \\ \hline
Picture resolution & 1 megapixel sufficient, higher is better  \\ \hline
Picture rate & 1 frame per second allows video  \\ \hline
Battery life & Depends on photo and flash rate, 2-5+ weeks typical on 6 C-cell batteries  \\ \hline
Trigger time & 2/10th second, longer will miss animals passing by  \\ \hline
Memory & 1gb compact flash cards \\ \hline 
Cost & $~\$500$ now, cheaper is better  \\ \hline
\end{tabular}
\label{tab:cam-req}
\end{table*}

We used Reconyx RC55 Camera traps. Figure~\ref{fig:bci-camera} shows a camera trap deployment at BCI.
 Compared to other models, these camera traps hold up well in the harsh rainforest conditions. However, any electronics will suffer under this humidity, so some care is needed. We keep small packets of desiccant in cameras, and regularly return them to the lab for cleaning and to dry them out in a dry-closet. Carefully designed enclosures are key in keeping the moisture away, and Reconyx cameras come with a custom plastic enclosure with a rubber gasket. Impact of environment on camera is presented in detail in Section~\ref{sec:results}. We use high-end rechargeable c-cells whose battery life depends on the level activity at a site especially the number of flash pictures. Experience from our deployments showed that we typically had an average of 30 days of battery life. 

\begin{figure*}
\begin{center}
\mbox{
\includegraphics[scale=0.45]{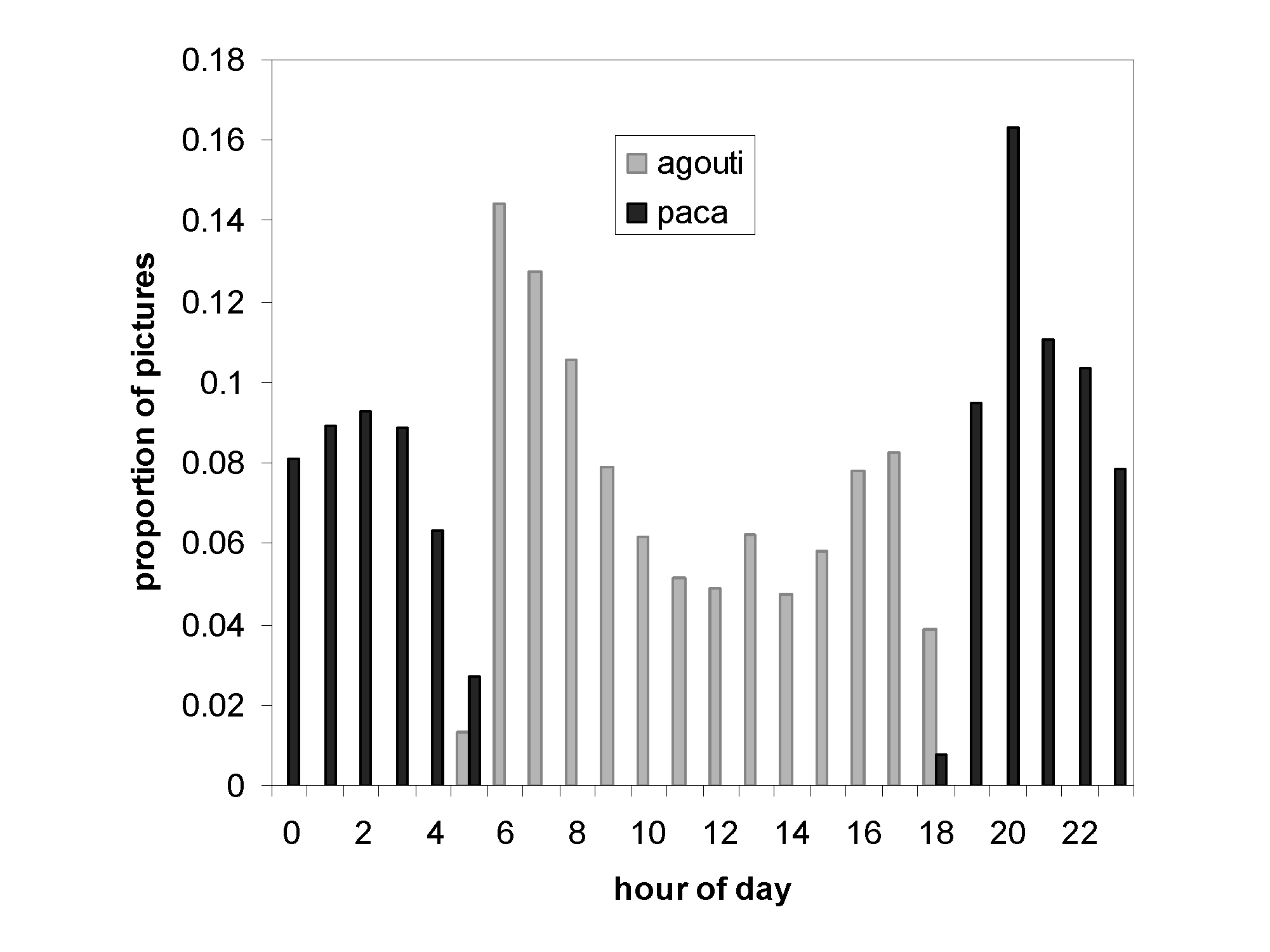}
}
\caption{The daily activity pattern for one nocturnal and one diurnal animal species for a BCI study.}
\label{fig:animal-activities}
\end{center}
\end{figure*}

We have a developed a simple data management infrastructure (based on MySQL open source database technology) that helps us manage these images efficiently (described in Section~\ref{sec:procedure}).


\section{Procedure}\label{sec:procedure} 
\subsection{Study Desgin}
The exact way in which cameras are placed in the landscape of a study area depend on the overall objectives.  A good review of this can be found in~\citep{Kays08, Long08}.


The number of cameras used in each sample unit represents a balance between collecting the best possible data and making the most efficient use of a limited number of cameras. Statistics are typically run on sample locations, so surveying more sites will give more statistical power. This is typically limited by the number of cameras owned by the study, transportation costs/time between camera sites, and the length of deployment for each camera. Deciding how long to deploy cameras at site reflects an important tradeoff between improving the likelihood of detecting a species at a given site (i.e., longer runs are better); increasing the number of different sample units that can be surveyed during the field season (shorter runs mean you can survey more sites); and, for some objectives, maintaining population closure at a site (i.e., no immigration or emigration). In addition to survey duration, the set type (e.g., baited or unbaited), the number of remote cameras within a sample unit, the geographic spread of the sample unit, and local animal density will also affect detection probability.

We assessed the mammal community with 20 Reconyx RC55 digital camera
traps with 1 Gb compact flash cards for image storage. Two cameras were
deployed simultaneously at random locations within each of 10 1-ha plots. 
To compare this randomized protocol with traditional trail-side sets we also deployed a subset of cameras along trails near our plots (Figure~\ref{fig:bci-biased-ct}).
We used a GPS unit (Garmin 60CSx) to locate
these points in the field and then mounted cameras on the nearest tree 
at a height of 20 cm (Figure~\ref{fig:bci-camera}). The camera view 
was maximized by aiming them in the most suitable
direction, with the least vegetation or slope obstructing their view within 5-10m. 
Cameras were programmed with the following settings: low-resolution (1 mega pixel) pictures at a frame rate of approximately 1 fps and trigger was set to no-delay mode.  They were also programmed to also make time-lapse pictures every 12 hours in order to check proper functioning. 

We scheduled camera deployments to be 8 days, whereupon the camera was moved to a new location.  Most analyses are done across camera sites, so decreasing the duration of each deployment to increase the number of sites surveyed is preferred.  However, increasing the number of days at a site will increase the precision of the estimate of passage rate. 
Areas with lower mammal density than BCI should use longer camera deployments, 2-4 week-long deployments are typical in other studies~\citep{Kays08}. This trade-off can be statistically modeled to help fine tune a study-design to meet specific research objectives~\citep{MacKenzie06}. 

We now briefly describe our data management infrastructure. After 8 days of deployment, we swap the memory cards in cameras with blank memory cards and return the used memory cards to the lab where images are  organized in a custom-made MySQL database with a PHP web interface. Time, date, trigger event, trigger type and camera are
automatically extracted from the metadata of the images (exif data).
Data is organized per plot, location, camera run and card run. Figure~\ref{fig:bci-dbschema} shows the the conceptual overview of the database but not the details. 

Because we record sequences of images consecutively (pseudo-video), we have to separate or join sequences before analyzing animal passage rates.  Our basic goal is to have each sequence represent one individual or social group of animals.  We consider any pair of sequences separated by more than 40 minutes to be different, and any less then 30 seconds apart to be the same.  Consecutive sequences with 
with intermediate interval lengths are flagged and checked manually to determine if they should be split or lumped. The final step of data processing is to identify the contents of each image sequence.  We register the species present and register the number of animals.  For species that are identifiable by unique coat patterns, such as ocelots or paca, we also note each animal's individual ID.
Figure~\ref{fig:bci-clip} shows a sample clip processed using the above procedure.

\section{Statistical Analysis}\label{sec:stat}

Camera trap data is analyzed in three main ways.  First, details of the animals represented by each photo sequence are available including the species, group size, date, time, and location.  This data is useful for showing the overall frequency of detection of each species and the temporal distribution of activity (Figure~\ref{fig:animal-activities} and Figure~\ref{fig:fs-sign}).  

Second, details of the animals detected at each camera location are calculated.  The simplest estimate is a detection rate over the entire camera deployment, recorded as the total number of sequences of a given species divided by the total time a camera was running.  This is useful as a general index of abundance that may also be used to estimate true animal density~\citep{Rowcliffe-08}.  A slightly more complicated query outputs the performance of each camera on each day it was in operation in terms of the detection or non-detection of a given species.  This data is analyzed to calculate the probability of detection for a given site, which can be further developed in occupancy modeling, taking into account various environmental covariates~\citep{MacKenzie06}.

Third, the captured histories of individual animals can be analyzed using mark-recapture protocols~\citep{MacKenzie06}. This is typically possible for a subset of species that have unique coat markings such as spotted jaguars or striped tigers~\citep{Karanth98}, but may also be applied to male ungulates with unique antler patterns~\citep{Jacobson97} or to other species tagged with unique color markings~\citep{Fuller96}.

\section{Experimental Results}\label{sec:results}

\subsection{Forest Signatures}

\begin{figure*}
\begin{center}
\mbox{
\subfigure[Frequency of detection for 25 species of terrestrial  birds and mammals on BCI. The horizontal dotted line at Y=1000 stands for the theoretical total maximum number of species at BCI. \label{fig:sigpolt1}]{\includegraphics[scale=0.4]{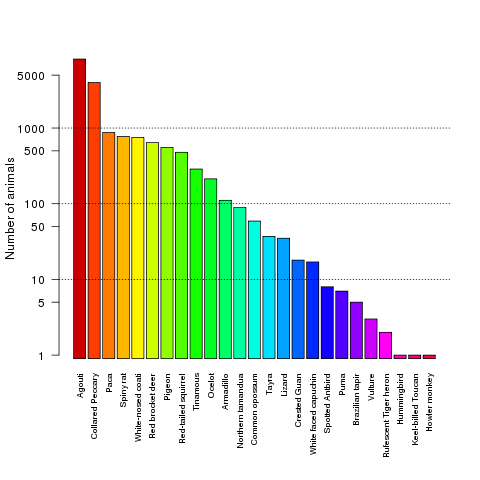}} \quad
\subfigure[The daily pattern of animal activity on the forest floor. Colors match the species names in Fig~\ref{fig:sigpolt1}.\label{fig:sigplot2}.]{\includegraphics[scale=0.45]{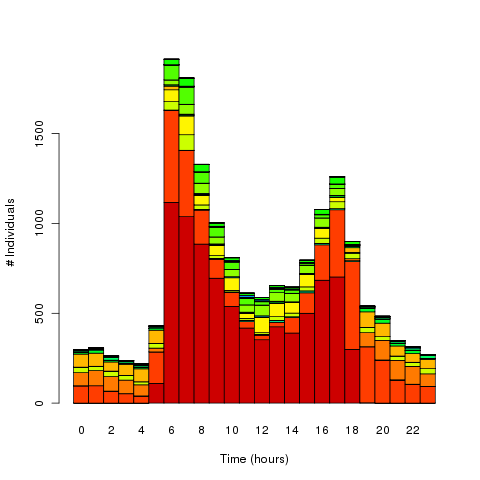}}
}
\caption{Forest signatures from camera trap deployment at  BCI}
\label{fig:fs-sign}
\end{center}
\end{figure*}    

Our year-long deployment of remote cameras at randomized locations has given us a unique and unbiased view of the overall activity of animals on the rainforest floor including the species present and their relative abundance (Figure~\ref{fig:fs-sign}). The deployment resulted in a total of 764 deployments with 17111 animal detections and 25 species detections. This measure of animal activity is simply the number of times a given species walked across a sample plot, and offers a direct metric of potential ecological impact. For example, as shown in Figure~\ref{fig:fs-sign}, on BCI agoutis, peccaries, and paca are the most frequently detected species, and thus the most likely to have an impact on local plant populations through seed predation or dispersal. If calibrated into density (animals/km$^2$)~\citep{Rowcliffe-08} these could also be used to derive estimates of biomass for each species or ecological group. 
	
The standardized measures of species diversity and abundance represented by these ÔsignaturesÕ (Figure~\ref{fig:fs-sign}) are exactly those needed to evaluate the effects of modern environmental change.  Effects of climate change and invasive species would be reflected in changes in species composition, while changes in abundance would reflect natural population fluctuations, as well as more dramatic crashes or explosions typical of human influenced dynamics.


\subsection{Sample Size Optimization}
\begin{figure*}
\begin{center}
\mbox{
\subfigure[Estimation of species diversity with increased sample size. Each camera deployment represents 8 days of monitoring.  Curves are drawn using a rarefaction (Sobs) or Jackknife (Jack1) resampling of 200 camera deployments on BCI.\label{fig:power-diversity}]{\includegraphics[scale=0.65]{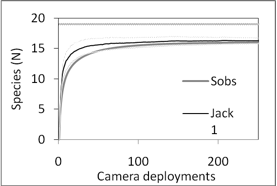}} \quad
\subfigure[The variation in estimated detection rate  for agoutis with sampling effort. The mean rate (black line) changes little, but the variation (min/max are thin red lines, 95\% confidence intervals are thick red lines) in estimates decreases with increasing sample effort, leveling off after around 15-20 camera deployments. Each camera deployment is 8-days long. All estimates come from 1000 resamples of data from one study plot.\label{fig:power-trap}]{\includegraphics[scale=0.40]{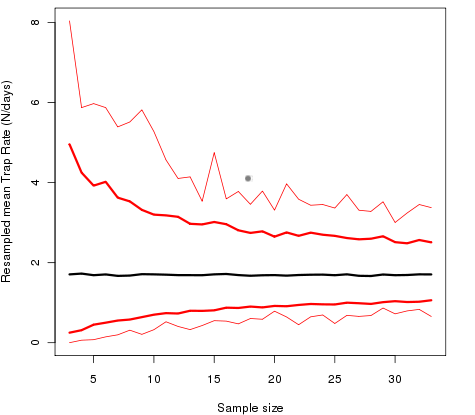}}
}
\caption{Sample Size Optimization Study Results}
\label{fig:power-study}
\end{center}
\end{figure*}    

Our year-round survey is unique in offering a seasonal view on the animal community.  However, many basic objectives of estimating the diversity and abundance of the community can be met with less effort.  We used our year-round data set to evaluate the sample size needed to meet these objectives.   
Figure~\ref{fig:power-diversity} shows the relationship between estimated mammalian species diversity and sampling effort. Each deployment represents one camera in the field for 8 days, and levels off after 15-25 deployments. There are 19 large and medium-sized terrestrial mammal species theoretically possible on BCI, although 4 of these (jaguar, jaguarundi, margray, and grison) are very rarely recorded on BCI, probably only as sporadic visitors. 

We also evaluated the sampling intensity needed to obtain an accurate estimate of detection frequency, an index of animal abundance (Figure~\ref{fig:power-trap}). This shows that the variation in average agouti detection rate levels off after 15-20 camera deployments, suggesting this is an appropriate sample effort. This could be met, for example, with 15 8-day deployments of one camera, or 3 deployments of 5 cameras. This relationship varies across species, with accurate estimates for species that are rare, or variable in their activity, requiring more sample effort.

\subsection{Camera Deployments Strategies}
\begin{figure*}
\includegraphics [scale=0.8]{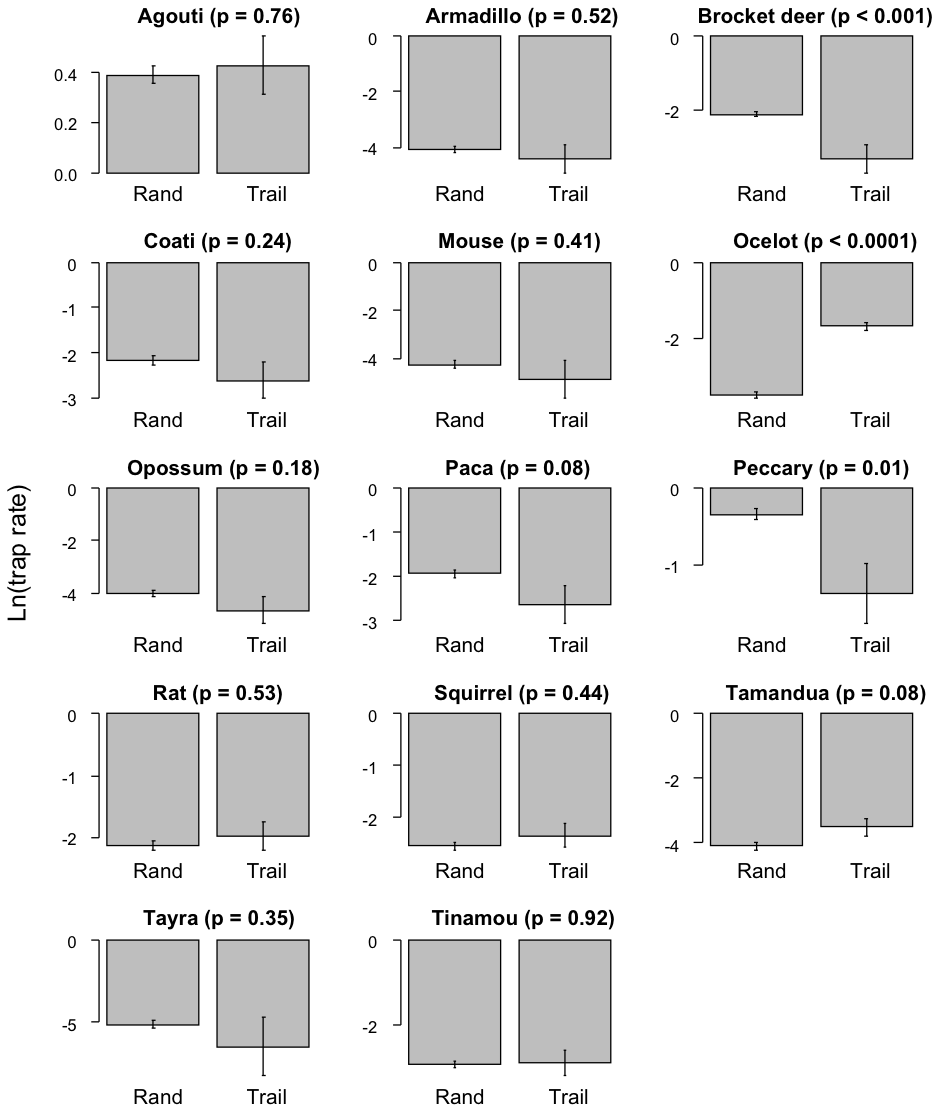}
\caption{Comparison of trap rates (log transformed) between random (Rand, n=905) and trail (Trail, n=76) camera placements for 14 species. Error bars are standard errors. P-values for each species give the significance of the difference between trail and random using F-tests (allowing for overdispersion) on quasi-Poisson generalised linear models of species counts, controlled for deployment duration by including the log of this value as an offset.}
\label{fig:deployment-strategies}
\end{figure*}

Sensor deployment and placement strategies has received considerable attention from the research community~\citep{Dhillon-03, Gonz-01,
Tilak-02, Wu-07, Xu-07}. However, to the best of our knowledge, this is is first study that takes into account application-level metric in a year-long real-world deployment. To evaluate the effect of camera placement on animal detections we compared the detection rate for cameras placed right on hiking trails (n=76) with those places in random locations within the forest (n=905).  
We found that there was a significant difference between trail and random trap rates for three out of 14 species tested (Figure~\ref{fig:deployment-strategies}). Ocelots favour trails (6-fold higher trap rate on trails), while brocket deer and peccary avoid them (respectively 3.3 and 2.8-fold higher trap rates on random placements). Paca also show a non-significant tendency to avoid trails, while tamandua show a slight tendency to favor them, but none of the other nine species show any evidence for a difference in trap rates between random and trail placements. Thus, in our study area, trail side cameras appear to be giving a biased view for a minority of species, although the degree of bias where it exists can be very high. It may also be worth noting that serious bias occurs only in the larger species ($>$10 kg) in this set.

\subsection{Spatial Autocorrelation Of Detection Rates:}


\subfiglabelskip=0pt
\begin{figure*}
\centering
\subfigure [ ] [ ] {%
\label{fig:rat_close}
\includegraphics[scale=0.55]{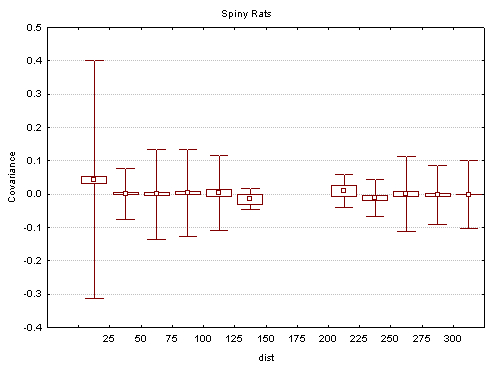}}
\hspace{8pt}%
\subfigure [ ] [ ] {%
\label{fig:agouti_close}
\includegraphics[scale=0.55]{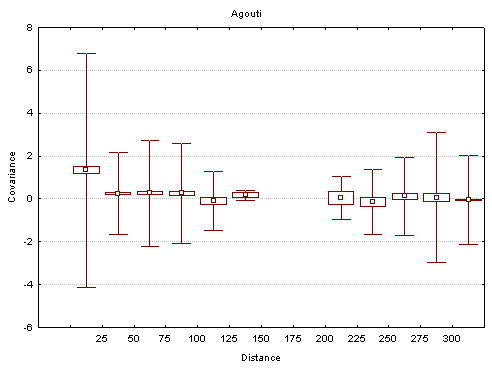}}
\subfigure [ ] [ ] {%
\label{fig:coati_close}
\includegraphics[scale=0.55]{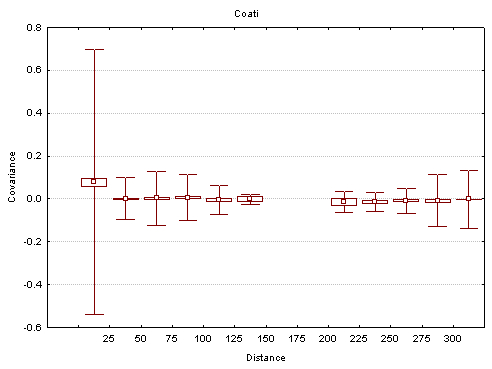}}
\hspace{8pt}%
\subfigure [ ] [ ] {%
\label{fig:deer_close}
\includegraphics[scale=0.55]{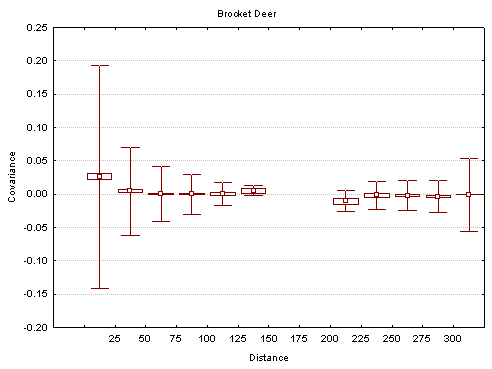}}
\subfigure [ ] [ ] {%
\label{fig:pec_close}
\includegraphics[scale=0.55]{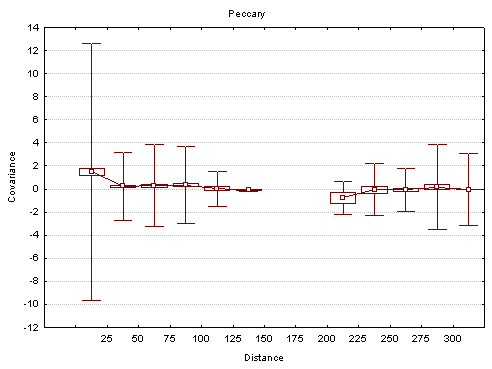}}
\hspace{8pt}%
\subfigure [ ] [ ] {%
\label{fig:peccary}
\includegraphics[scale=0.55]{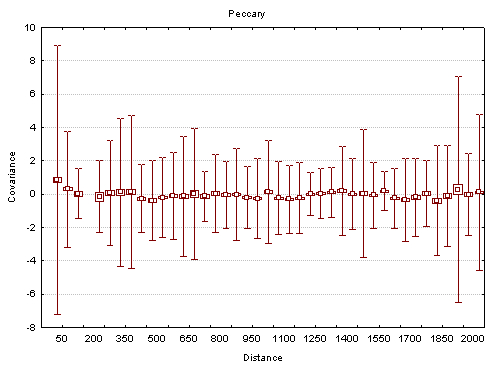}}

\caption{Semivariograms of detection rates for five species of mammals recorded by camera traps showing the decline in spatial autocorrelation after 25m out to 300m \subref{fig:rat_close} Spiny Rat  \subref{fig:agouti_close} Agouti, \subref{fig:coati_close} Coati, \subref{fig:deer_close} Brocket Deer, \subref{fig:pec_close} Peccary. Test for autocorrelation at larger scales produced similar results for all species, as represented by one graph for our largest species \subref{fig:peccary}. Graphs show the mean (center square), standard error (box) and standard deviation (wisker) for all pairs of camera traps within a given distance class.
}
\label{fig:corr-mammals}
\end{figure*}

A common concern for all camera trap surveys not using mark/recapture analyses is to determine how far apart to camera sites must be to be spatially independent.  Typically, studies take extreme caution in this regard, spacing camera traps far enough apart to minimize the potential that the same individual animal would be detected by two cameras.  The usual measure is to estimate the diameter of the home range of a target species and make this the minimum spacing for cameras, which is often many km~\citep{Gompper-06}.  However, no study has empirically evaluated the autocorrelation of camera trap data.  Our data presents an excellent opportunity to do this, with pairs of cameras within a plot offering small-scale comparisons, and comparisons across plots offering larger-scale comparisons.
	We used the Geostatistical analyst extension of ArcGIS9 (ESRI) function to evaluate spatial autocorrelation by comparing the detection rate for a given species across all pairs of cameras.  We analyzed data within 2-month time windows to take into account that spatial patterns of animal activity may vary seasonally. We analyzed data for the five most common species, which include a large range of body size and scale of movement across the landscape. 
 
	The main result was that we found very little spatial autocorrelation in animal detection rates for any of the $5$ mammal species considered.  Figure~\ref{fig:corr-mammals} shows this result in that the covariance (y axis) between all pairs of traps is not significantly different from 0 when the cameras are greater than 25m apart (x axis).  The results were similar across species, with a positive correlation between the detection rates for camera sites very close to each other ($ < 25$ m) but not at any other spatial scales. This result suggests that cameras can be placed much closer to each other than is typically done and still record statistically independent data; instead of the many km minimum distance, we suggest cameras be a minimum of 25 m apart. 
Cameras set more closely might still be useful, if spatial autocorrelation is irrelevant, or taken into account by analyses.


\subsection{Camera Performance In Real-world}

\begin{figure*}
\begin{center}
\mbox{
\subfigure[Impact of seasonality on camera performance.\label{fig:ct-perf}]{\includegraphics[scale=0.45]{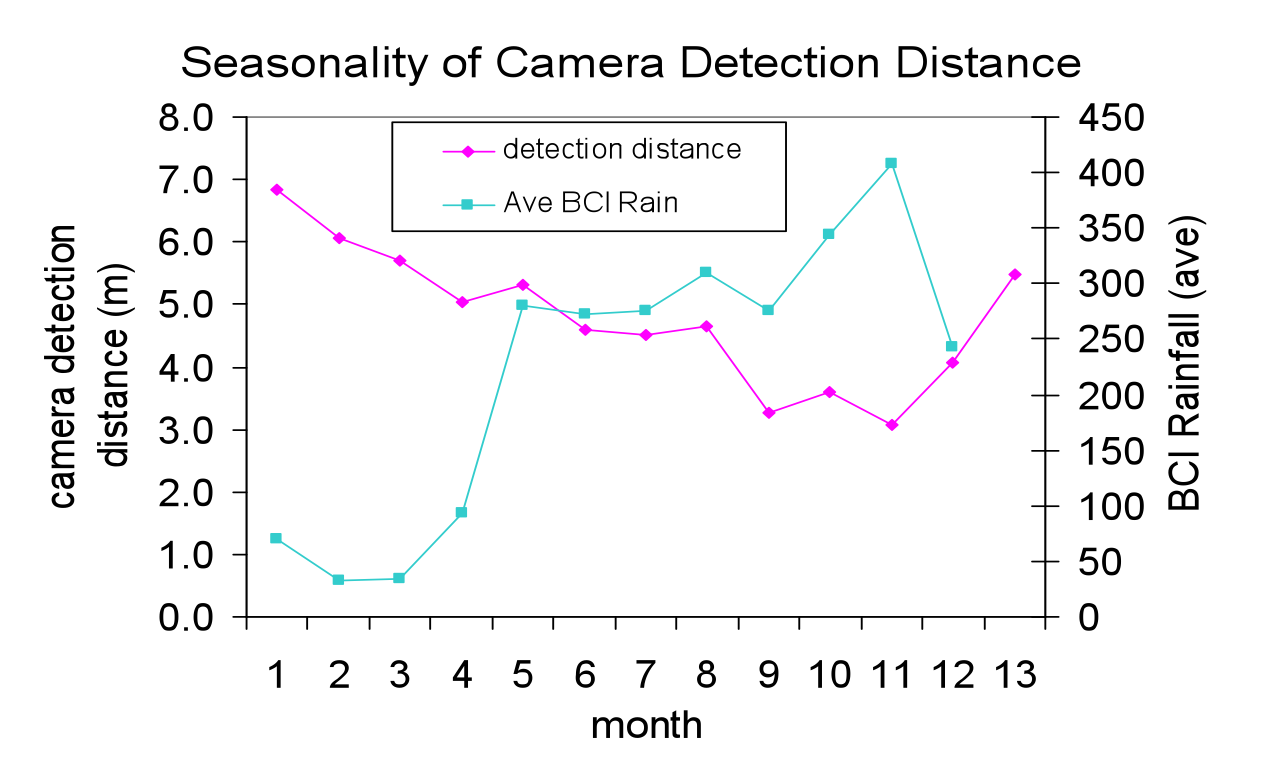}} \quad
\subfigure[Camera failures in our study.\label{fig:camera-failures}]{\includegraphics[scale=0.4]{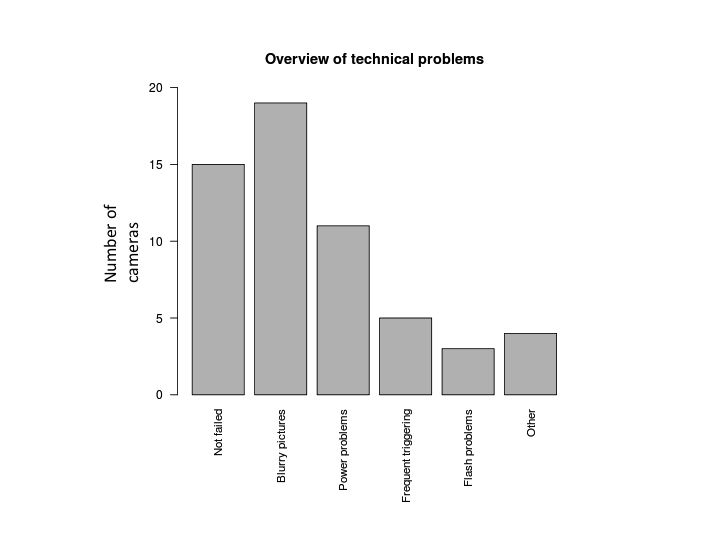}}
}
\caption{Study of Camera Performance in Real-World}
\label{fig:ct-rw}
\end{center}
\end{figure*}    

Due to challenging weather and environmental conditions camera traps are often more difficult to operate in rainy reasons. To minimize the impact of seasonality on camera performance we suggest  use of silica desiccant packets (2 if possible) to keep the insides dry. 




There was a strong effect of seasonality Figure~\ref{fig:ct-perf}, with detection distance (measured by walking in front of the camera when setting it out) shortening during the rainy season.  This is probably a combination of moisture on the sensor, in the air between the sensor and the target, and on the target itself.  Together, these would dampen the difference between the IR signature of an animal compared with the background, and thus reduce its ability to detect an animal.
Shrinking the effective area each camera surveys has obvious impacts on the number of animals it detects.  Thus it is important to document these effects, and take them into account for comparisons of animal activity across seasons or sites. 
We also advise keeping cameras in dry-closet whenever not in use. 
Based on our experience rotating cameras out of service every 2 months for preventative maintenance works well. 

\subsection{Relation Between Animal Size And Camera Parameters}

\begin{figure*}
\includegraphics [scale=0.6]{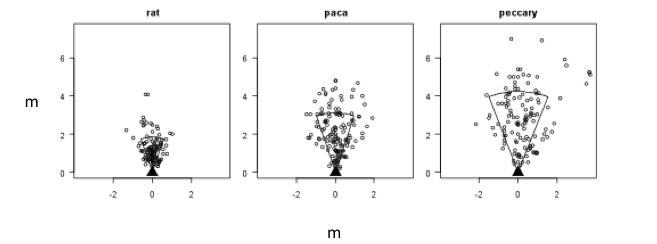}
\caption{Positions of animals on first detection (open points) relative to the camera (filled triangles) for three representative species of contrasting size: spiny rat (0.4 kg), paca (8 kg) and peccary (25 kg). The camera is at the origin, with axis values in metres. The open sectors illustrate effective detection zones, estimated by fitting detection models to the distance and angle data for locations.}
\label{fig:detect-dist-angle}
\end{figure*}

An inherent property of the triggers used on camera traps is that they are less likely to detect small animals than large, all else being equal. This can be seen in the distribution of positions of different species relative to the camera when first detected (Figure~\ref{fig:detect-dist-angle}), demonstrating much shorter average distances for the smaller species and, to a lesser extent, narrower angles. We are currently developing methods to model this phenomenon~\citep{Rowcliffe-sub10}, allowing us to quantify camera sensitivity for any given species, camera or environment (illustrated by detection zone sectors in (ref. Figure~\ref{fig:detect-dist-angle}). This approach will be important in enabling us to extract abundance signals from trap rates by controlling for camera sensitivity.


\subsection{Camera Failures In Real-world}

We observed that only 30\% of the deployed cameras never failed during the year (Figure~\ref{fig:camera-failures}). This shows that operating a camera trap based solution over extended periods of time does require monitoring and debugging. Approximately 40\% failures  were due camera lens being blurry. 
The manufacturer repaired all cameras, and used our experience to find that the problem was caused by humidity de-laminating a filter on the lens.  They have since improved the seal on the lens.  
The second major source of failures was (-20\%) caused by humidity affecting circuitry of the camera.  The manufacture has since developed a new coating for their circuit boards which should improve their performance in high humidity.

\section{Practical Considerations and Experiences}\label{sec:pract}


\subsection{Equipment Management:} Although we are hopeful that improved designs will be more weather-proof, we expect that hardware maintenance will remain a critical aspect for any long-term monitoring project. Key items to regularly inspect and service include the rubber gaskets that prevent moisture
from entering system components, exposed metal contacts and battery leads (for corrosion and dirt),
and external wires. In humid environments, care should be taken when moving cameras from air-conditioned rooms into field conditions because condensation will form on electronic parts and lenses. Finally, it is important to be realistic about the durability of remote cameras and to prepare for
equipment malfunction. We advise that researchers not deploy every available remote camera but rather
have a few extra units at the ready to replace broken equipment. When working in particularly challenging environments, maintaining a reserve of cameras amounting to 20\% of the total number deployed may be necessary to maintain consistent sampling effort.

\subsection{In-Field Equipment Checking}
Determining the optimal frequency at which to check each camera station usually entails a tradeoff between maximizing efficiency and ensuring that stations will remain functional during the entire sampling interval.  Long deployments record more animals, but may run out of batteries or memory, and result in fewer sites being sampled with a limited number of cameras. Numerous factors, including the type of camera system (e.g., film versus digital), camera programming (e.g., camera delay), site remoteness, whether sets are baited or unbaited, and expected site activity level, must be considered.  When surveying a new site we generally recommend initially checking camera sets within 7-14 days.  From these preliminary data you can judge optimal survey length based on remaining battery/memory and the rate of detection of target species.



 \subsection{Baiting and Site Selection}
The study design used to deploy camera traps is dependent on the goal. When wanting to determine absence or presence of a target species baiting cameras can be considered. Baiting will attract animals from a wider area, but will only attract target species and may even repel other species. Another way of increasing trapping rate is using landscape feature that animals use. These can be trail (human or game) but also for example drinking ponds. To get an unbiased sample, random camera locations can be considered. By generating random locations and deploying cameras at these sites one tries to prevent having biases in the animals sampled.

 \subsection{Labor Estimates}
A single experienced person can efficiently set up, check, and break down most remote camera systems. Given the logistics and challenges of most surveys, however, including weather, accessing the site (e.g., by vehicle, on foot), safety, the number of stations, and the total amount of weight to be carried into the field - most surveys use two-person field teams. Furthermore, if inexperienced personnel or new equipment are being used, two-person teams will likely improve the success of the effort and provide more learning opportunities.

\subsection{Minimizing Theft:} In this project we had minimal risk of theft because of the high security on BCI.  However, this is a potential problem for many distributed sensor networks, including camera traps.  Minimizing the detection of your camera by others is the first measure to take to reduce theft.  Running cameras off-trail and below eye-level helps this.  A visible flash also gives away the location of your camera, so digital cameras with IR flashes should be more cryptic.  Units with a red filter over the flash further reduce risk by eliminating even the dull red glow of IR flashes.  A simple cable and padlock should deter most thieves. A small sign taped to the side of the camera briefly describing the purpose of the study and providing relevant contact information typically satisfies curiosity and limits vandalism.  However, no lock is foolproof to a determined thief with the right tools, so studies should anticipate some level of theft by having replacement cameras on hand.  

\section{Future Work}\label{sec:futurework}

\subsection{Automated Image Analysis Framework}

\begin{figure*}
\begin{center}
\mbox{
\subfigure[Example calibration target.\label{fig:camera-calib}]{\includegraphics[scale=0.85]{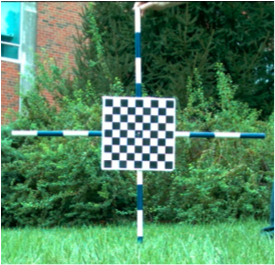}} \quad
\subfigure[Camera physical parameter estimation.\label{fig:camera-paramest}]{\includegraphics[scale=0.85]{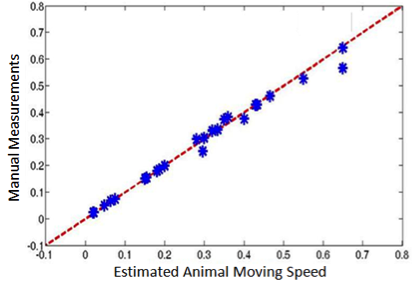}}
}
\caption{Forest signatures from camera trap deployment at  BCI}
\label{fig:img-analysis}
\end{center}
\end{figure*}

The hardware and data management protocols we outline here are appropriate for the research objectives of most typical camera-trap projects, and they can easily handle the tens of thousands pictures we generate per year. However, expanding collaborative wildlife monitoring networks with large spatiotemporal scales will cause new data analysis and management problems. 

Camera traps are inherently fixed in space, and moving animals are the primary cause of content changes between video frames. We have been exploring the application of automated and semi-automated image content analysis techniques to camera trap images. We believe that these capabilities have the potential to greatly improve the efficiency of large scale camera-trap image processing and enable practical extraction of new information from camera trap images that would support the investigation of a number of interesting scientific questions.  For example, automated identification of the animal species would be of obvious utility. Even if the identification is not 100\% accurate, pre-sorting images by species before presenting them to experts for analysis could make the species identification process significantly more efficient.  Likewise, advanced pattern recognition algorithms could help researchers to efficiently identify individual animals in those species with unique coat patterns (e.g. tigers or jaguars).
Furthermore, after camera calibration using a pattern as shown in Figure~\ref{fig:camera-calib}, animal movements and size can be extracted from the camera trap images. We can then estimate the position of the animal relative to the camera at the moment it triggered the motion sensor, and measure the distance to, and speed of, the animal as it moves across the field of view. The position of the animal at triggering is important because it can be used to determine the camera detection footprint for different species and types of sites~\citep{Rowcliffe-08}. Measurement of the speed an animal move might is also useful to calibrate estimates of animal density from camera trap detection.

While these two measurements can, and have, been made manually in the field, they are quite time consuming.  We now describe our ongoing effort on automated image processing for camera trap videos for animal detection, location and speed estimation. 

Estimating Physical and Motion Parameters from each camera trap image sequence where an animal is detected, we estimate its motion and physical parameters, such as average and maximum moving speeds, entry position/angle, and body size using the following procedure. 

The first and enabling step is camera calibration~\citep{Hartley-04, Zhang-00}. To minimize the workload of people deploying the cameras, we choose a self-calibration approach~\citep{Maybank-92, Huang-04} using a known calibration object. For example, we used a one meter long striped white stick placed at different orientations during our camera-trap data collection at BCI. A checkerboard pattern is also often used. Figure~\ref{fig:camera-calib} shows a general purpose calibration target that we have used for some of our experiments. From multiple images of the calibration target placed at a few locations in the field of view of the camera we derive calibration parameters for both the camera and the site.

Parameters specific to the camera are known as ÒintrinsicÓ parameters and those associated with the placement of the camera relative to the scene are called ÒextrinsicÓ parameters. It is possible to estimate both the intrinsic and extrinsic parameters from the field images if sufficient calibration images are collected. However, this approach places a somewhat significant burden on the field crew. So, our preferred approach is to separately perform the intrinsic calibration on each camera prior to deployment in the field. Then an extrinsic calibration can be done for each site with significantly reduced effort in the field.

 Intrinsic and extrinsic camera parameters are determined by building a camera model that produces the minimum projection error~\citep{Hartley-04}.  From the model we can then estimate the physical location (on the ground plane) and the dimension of the animal. 
 
 Currently we use a semi-automatic point tracking approach with some manual point designation to derive estimates of the animalÕs movement. 
In our preliminary tests with BCI and other data we have found this approach to provide accurate estimates of animal movement speed, and the distance from the camera at first detection. Figure~\ref{fig:camera-paramest} shows estimated animal moving speed from camera-trap data in comparison with the manual measurements taken in the field (a laborious process).  We can see that our estimates from camera trap images match the manual measurements very well.

In future work, we will explore an alternate approach that would be integrated with an automatic animal recognition process. In this approach we would treat this problem as a 3-D camera view geometry problem: determining a minimum vertical rectangular bounding box B of the animal in the 3-D space whose bottom edge lies on the ground plane and its projection A in the image plane contains the extracted animal silhouette. The center and dimensions (height and width) of box B are then the estimated body position and dimensions of the animal. 
With the animal body position in each image available, we can then track the body position over time and determine the animalÕs instant moving speed, moving trajectory, maximum and average moving speed, principle moving direction, average stopping frequency, distance and angle to first detection, entry angle, and other motion-related parameters. The team is working to determine additional useful motion and physical parameters that might be derived from this data. 

\subsection{Networking Cameras}

We do not network our cameras or retrieve data in real time. The energy budget needed to transmit so many images would require a much larger battery or solar panel, which are not practical or our applications. Real-time data would be useful for responding to rare events or monitoring camera performance.  Furthermore, it would make very difficult camera sites more practical to monitor (e.g. treetop canopies) without physically visiting the site to retrieve the images. However, in future, we plan to explore the option of real-time data transmission by networking the field deployed cameras. This will allow us to study various interesting system level issues such as data transmission reliability and energy consumption.

\section{Conclusion}\label{sec:conclusion}
To conclude, data gathered from our a year-long terrestrial animal monitoring system at Barro Colorado Island, Panama shows the spatio-temporal dynamics of terrestrial bird and mammal activity at the site - data relevant to immediate science questions, and long-term conservation issues. We believe that the experience gained and lessons learned during our year long deployment and testing of the camera traps are applicable to broader sensor network applications and are valuable for the advancement of the sensor network research.
\biboptions{longnamesfirst}
\bibliographystyle{elsarticle-num}
\bibliography{movebank,wsn}

\begin{thebibliography}{10}
\expandafter\ifx\csname url\endcsname\relax
  \def\url#1{\texttt{#1}}\fi
\expandafter\ifx\csname urlprefix\endcsname\relax\def\urlprefix{URL }\fi
\expandafter\ifx\csname href\endcsname\relax
  \def\href#1#2{#2} \def\path#1{#1}\fi

\bibitem{Cerpa01}
A.~Cerpa, J.~Elson, M.~Hamilton, J.~Zhao, D.~Estrin, L.~Girod, Habitat
  monitoring: application driver for wireless communications technology, in:
  SIGCOMM LA '01: Workshop on Data communication in Latin America and the
  Caribbean, 2001.

\bibitem{Dinh07}
T.~L. Dinh, W.~Hu, P.~Sikka, P.~Corke, L.~Overs, S.~Brosnan, Design and
  deployment of a remote robust sensor network: Experiences from an outdoor
  water quality monitoring network, in: Proceedings of the 32nd IEEE Conference
  on Local Computer Networks, 2007.

\bibitem{Gilman05}
G.~Tolle, J.~Polastre, R.~Szewczyk, D.~Culler, N.~Turner, K.~Tu, S.~Burgess,
  T.~Dawson, P.~Buonadonna, D.~Gay, W.~Hong, A macroscope in the redwoods, in:
  SenSys '05: Proceedings of the 3rd international conference on Embedded
  networked sensor systems, 2005.

\bibitem{Guillermo08}
G.~Barrenetxea, F.~Ingelrest, G.~Schaefer, M.~Vetterli, O.~Couach, M.~Parlange,
  Sensorscope: Out-of-the-box environmental monitoring, in: IPSN '08:
  Proceedings of the 7th international conference on Information processing in
  sensor networks, 2008.

\bibitem{Lynette06}
L.~Laffea, R.~Monson, R.~Han, R.~Manning, A.~Glasser, S.~Oncley, J.~Sun,
  S.~Burns, S.~Semmer, J.~Militzer, Comprehensive monitoring of co2
  sequestration in subalpine forest ecosystems and its relation to global
  warming, in: SenSys '06: Proceedings of the 4th international conference on
  Embedded networked sensor systems, 2006.

\bibitem{Mainwaring02}
A.~Mainwaring, D.~Culler, J.~Polastre, R.~Szewczyk, J.~Anderson, Wireless
  sensor networks for habitat monitoring, in: WSNA '02: Proceedings of the 1st
  ACM international workshop on Wireless sensor networks and applications,
  2002.

\bibitem{Selavo07}
L.~Selavo, A.~Wood, Q.~Cao, T.~Sookoor, H.~Liu, A.~Srinivasan, Y.~Wu, W.~Kang,
  J.~Stankovic, D.~Young, J.~Porter, Luster: wireless sensor network for
  environmental research, in: SenSys '07: Proceedings of the 5th international
  conference on Embedded networked sensor systems, 2007.

\bibitem{Szewczyk04}
R.~Szewczyk, A.~Mainwaring, J.~Polastre, J.~Anderson, D.~Culler, An analysis of
  a large scale habitat monitoring application, in: SenSys '04: Proceedings of
  the 2nd international conference on Embedded networked sensor systems, 2004.

\bibitem{Szlaveczcorr07}
K.~Szlavecz, A.~Terzis, S.~Ozer, R.~Musaloiu-Elefteri, J.~Cogan, S.~Small,
  R.~C. Burns, J.~Gray, A.~S. Szalay, Life under your feet: An end-to-end soil
  ecology sensor network, database, web server, and analysis service.

\bibitem{Holger-07}
H.~Karl, A.~Will, Protocols and Architectures for Wireless Sensor Networks,
  Wiley-Interscience, 2007.

\bibitem{Kahn-99}
J.~M. Kahn, R.~H. Katz, K.~S.~J. Pister, Next century challenges: Mobile
  networking for "smart dust", in: MOBICOM, 1999, pp. 271--278.

\bibitem{Krishnamachari-05}
B.~Krishnamachari, Networking Wireless Sensors, Cambridge University Press,
  2006.

\bibitem{Nathan-09}
R.~Nathan, W.~Getz, E.~Revilla, M.~Holyoak, R.~Kadmon, D.~Saltz, P.~Smouse, A
  movement ecology paradigm for unifying organismal movement research, PNAS
  (105) (2009) 19052--19059.

\bibitem{Turchin98}
P.~Turchin, Quantitative analysis of movement: measuring and modeling
  population redistribution in animals and plants, Sinauer Associates, Inc.,
  1998.

\bibitem{Lord62}
R.~Lord, F.~Bellrose, W.~Cochran, Radio telemetry of the respiration of a
  flying duck, Science 137 (1962) 39--40.

\bibitem{Kays08}
R.~Kays, K.~Slauson., Remote cameras. Noninvasive Survey Methods for North
  American Carnivores, Editors: R. Long and P. MacKay and J. Ray and W.
  Zielinski, Island Press, 2008.

\bibitem{MacKenzie-02}
D.~MacKenzie, J.~Nichols, G.~Lachman, S.~Droege, J.~Royle, C.~Langtimm,
  Estimating site occupancy rates when detection probabilities are less than
  one, Ecology (83) (2002) 2248--2255.

\bibitem{Pitra06}
C.~Pitra, P.~VazPinto, B.~OÕKeeffe, S.~Willows-Munro, B.~Jansen~van Vuuren,
  T.~Robinson, \href{http://dx.doi.org/10.1007/s10344-005-0026-y}{Dna-led
  rediscovery of the giant sable antelope in angola}, European Journal of
  Wildlife Research 52 (2006) 145--152, 10.1007/s10344-005-0026-y.
\newline\urlprefix\url{http://dx.doi.org/10.1007/s10344-005-0026-y}

\bibitem{Cardoza02}
J.~Cardoza, S.~Langlois, The eastern cougar: a management failure?, in:
  Wildlife Society Bulletin, Vol. (30), 2002, pp. 265--273.

\bibitem{Zielinski05}
W.~Zielinski, R.~Truex, F.~Schlexer, L.~Campbell, C.Carroll, Historical and
  contemporary distributions of carnivores in forests of the sierra nevada,
  california, usa, Journal of Biogeography (32) (2005) 1385--1407.

\bibitem{Rowcliffe-08}
J.~Rowcliffe, J.~Field, S.~Turvey, C.~Carbone, Estimating animal density using
  camera traps without the need for individual recognition, Journal of Applied
  Ecology (45) (2008) 1228--1236.

\bibitem{Karanth98}
K.~Karanth, J.~Nichols, Estimation of tiger densities in india using
  photographic captures and recaptures, Ecology 79(8) (1998) 2852--2862.

\bibitem{Karanth04}
K.~Karanth, J.~Nichols, N.~Kumar, W.~Link, J.~Hines, Tigers and their prey:
  predicting carnivore densities from prey abundance, Vol. 101, Proceedings of
  the National Academy of Sciences, 2004.

\bibitem{Karanth06}
K.~Karanth, J.~Nichols, N.~Kumar, J.~Hines, Assessing tiger population
  dynamimcs using photographic capture-recapture sampling, Ecology 87 (2006)
  2925--2937.

\bibitem{Jansen08}
P.~Jansen, S.~Bohlman, C.~Garzon-Lopez, Muller-Landau, S.~Wright, Large-scale
  spatial variation in palm fruit abundance across a tropical moist forest
  estimated from high-resolution aerial photographs, Ecography 31 (2008)
  33--42.

\bibitem{Hartung06}
C.~Hartung, R.~Han, C.~Seielstad, S.~Holbrook, Firewxnet: a multi-tiered
  portable wireless system for monitoring weather conditions in wildland fire
  environments, in: MobiSys '06: Proceedings of the 4th international
  conference on Mobile systems, applications and services, 2006.

\bibitem{Leigh99}
E.~Leigh, Tropical forest ecology: a view from barro colorado island, in:
  Oxford University Press, 1999.

\bibitem{Wegge04}
P.~Wegge, C.~Pokheral, S.~Jnawali, Effects of trapping effort and trap shyness
  on estimates of tiger abundance from camera trap studies, Animal Conservation
  7(3) (2004) 251--256.

\bibitem{Long08}
R.~Long, P.~MacKay, W.~Zielinski, J.~Ray, Noninvasive survey methods for
  Carnivores, Island Press, 2008.

\bibitem{MacKenzie06}
D.~MacKenzie, J.~Nichols, J.~Royle, K.~Pollock, L.~Bailey, J.~Hines, Occupancy
  estimation and modeling: inferring patterns and dynamics of species
  occurrence, Academic Press, 2006.

\bibitem{Jacobson97}
H.~Jacobson, J.~Kroll, R.~Browning, B.~Koerth, M.~Conway,
  \href{http://www.jstor.org/stable/3783491}{Infrared-triggered cameras for
  censusing white-tailed deer}, Wildlife Society Bulletin 25(2) (1997)
  557--562.
\newline\urlprefix\url{http://www.jstor.org/stable/3783491}

\bibitem{Fuller96}
T.~Fuller, E.~York, S.~Powell, T.~Decker, R.~DeGraaf, An evaluation of
  territory mapping to estimate fisher density, in: Canadian Journal of
  Zoology, Vol. 79(9), 2001, pp. 1691--1696.

\bibitem{Dhillon-03}
S.~Dhillon, K.~Chakrabarty, Sensor placement for effective coverage and
  surveillance in distributed sensor networks, Vol.~3, 2003, pp. 1609 --1614
  vol.3.
\newblock \href {http://dx.doi.org/10.1109/WCNC.2003.1200627}
  {\path{doi:10.1109/WCNC.2003.1200627}}.

\bibitem{Gonz-01}
H.~Gonz\'{a}lez-Banos, A randomized art-gallery algorithm for sensor placement,
  in: SCG '01: Proceedings of the seventeenth annual symposium on Computational
  geometry, ACM, New York, NY, USA, 2001, pp. 232--240.
\newblock \href {http://dx.doi.org/http://doi.acm.org/10.1145/378583.378674}
  {\path{doi:http://doi.acm.org/10.1145/378583.378674}}.

\bibitem{Tilak-02}
S.~Tilak, N.~B. Abu-Ghazaleh, W.~Heinzelman, Infrastructure tradeoffs for
  sensor networks, in: WSNA '02: Proceedings of the 1st ACM international
  workshop on Wireless sensor networks and applications, ACM, New York, NY,
  USA, 2002, pp. 49--58.
\newblock \href {http://dx.doi.org/http://doi.acm.org/10.1145/570738.570746}
  {\path{doi:http://doi.acm.org/10.1145/570738.570746}}.

\bibitem{Wu-07}
Q.~Wu, N.~S.~V. Rao, X.~Du, S.~S. Iyengar, V.~K. Vaishnavi, On efficient
  deployment of sensors on planar grid, Comput. Commun. 30~(14-15) (2007)
  2721--2734.
\newblock \href
  {http://dx.doi.org/http://dx.doi.org/10.1016/j.comcom.2007.05.012}
  {\path{doi:http://dx.doi.org/10.1016/j.comcom.2007.05.012}}.

\bibitem{Xu-07}
X.~Xu, S.~Sahni, Approximation algorithms for sensor deployment, IEEE Trans.
  Comput. 56~(12) (2007) 1681--1695.
\newblock \href {http://dx.doi.org/http://dx.doi.org/10.1109/TC.2007.1063}
  {\path{doi:http://dx.doi.org/10.1109/TC.2007.1063}}.

\bibitem{Gompper-06}
M.~Gompper, R.~Kays, J.~Ray, S.~LaPoint, D.~Bogan, J.~Cryan, A comparison of
  non-invasive techniques to survey carnivore communities in northeastern north
  america, in: Wildlife Society Bulletin, Vol.~34, 2006, pp. 1142--1151.

\bibitem{Rowcliffe-sub10}
J.~Rowcliffe, C.~Carbone, P.~A. Jansen, R.~K. B., Kranstauber, Quantifying the
  sensitivity of camera traps: an adapted distance sampling approach, Methods
  in Ecology and Evolution (submitted).

\bibitem{Hartley-04}
R.~I. Hartley, A.~Zisserman, Multiple view geometry in computer vision, in: 2nd
  Ed., Cambridge University Press, 2004.

\bibitem{Zhang-00}
Z.~Zhang, A flexible new technique for camera calibration, IEEE Transactions on
  Pattern Analysis and Machine Intelligence 22(11) (2000) 1330--1334.

\bibitem{Maybank-92}
S.~J. Maybank, O.~D. Faugeras, A theory of self-calibration of a moving camera,
  The International Journal of Computer Vision 8(2) (1992) 123Ð152.

\bibitem{Huang-04}
C.-R. Huang, C.-S. Chen, P.-C. Chung, An improved algorithm for two-image
  camera self-calibration and euclidean structure recovery using absolute
  quadric, in: Pattern Recognition, Vol. 37(8), 2004, pp. 1713--1722.

\end{thebibliography}
\end{document}